\documentclass[10pt,twocolumn,english,prd,superscriptaddress,nofootinbib,preprintnumbers,showpacs,floatfix]{revtex4-1}
\usepackage[utf8]{inputenc}
\usepackage{amsmath}
\usepackage{amsfonts}
\usepackage{amssymb}
\usepackage{graphicx}
\graphicspath{ {figures/} }
\usepackage[usenames,dvipsnames]{xcolor}
\usepackage{hyperref}   
\hypersetup{colorlinks=true, 
citecolor=MidnightBlue, linkcolor=MidnightBlue, urlcolor=Cyan}
\usepackage{tikz}
\usepackage{verbatim} 
\definecolor{purple}{rgb}{1,0,1}
\definecolor{lime}{HTML}{A6CE39} 

\newcommand{\orcidicon}{%
	\begin{tikzpicture}
	\draw[lime, fill=lime] (0,0) 
		circle [radius=0.16] 
		node[white] {{\fontfamily{qag}\selectfont \tiny ID}};
	\draw[white, fill=white] (-0.0625,0.095) 
		circle [radius=0.007];
	\end{tikzpicture}	\hspace{-2mm}
}
\newcommand\orcidFrancisco{{\href{https://orcid.org/0000-0002-9388-8373}{\orcidicon}}}
\newcommand\orcidManuel{{\href{https://orcid.org/0000-0001-8586-0285}{\orcidicon}}}
\newcommand\orcidTarciso{{\href{https://orcid.org/0009-0007-0450-2672}{\orcidicon}}}
\begin{document}
\title{(Regular) Black holes in conformal Killing gravity coupled to nonlinear electrodynamics and scalar fields}

	\author{Jos\'{e} Tarciso S. S. Junior\orcidTarciso\!\!}
 \email{tarcisojunior17@gmail.com}
\affiliation{Faculdade de F\'{\i}sica, Programa de P\'{o}s-Gradua\c{c}\~{a}o em 
F\'isica, Universidade Federal do 
 Par\'{a},  66075-110, Bel\'{e}m, Par\'{a}, Brazil}

	\author{Francisco S. N. Lobo\orcidFrancisco\!\!} \email{fslobo@fc.ul.pt}
\affiliation{Instituto de Astrof\'{i}sica e Ci\^{e}ncias do Espa\c{c}o, Faculdade de Ci\^{e}ncias da Universidade de Lisboa, Edifício C8, Campo Grande, P-1749-016 Lisbon, Portugal}
\affiliation{Departamento de F\'{i}sica, Faculdade de Ci\^{e}ncias da Universidade de Lisboa, Edif\'{i}cio C8, Campo Grande, P-1749-016 Lisbon, Portugal}

	\author{Manuel E. Rodrigues\orcidManuel\!\!}
	\email{esialg@gmail.com}
	\affiliation{Faculdade de F\'{\i}sica, Programa de P\'{o}s-Gradua\c{c}\~{a}o em 
F\'isica, Universidade Federal do 
 Par\'{a},  66075-110, Bel\'{e}m, Par\'{a}, Brazil}
\affiliation{Faculdade de Ci\^{e}ncias Exatas e Tecnologia, 
Universidade Federal do Par\'{a}\\
Campus Universit\'{a}rio de Abaetetuba, 68440-000, Abaetetuba, Par\'{a}, 
Brazil}

\date{\LaTeX-ed \today}
\begin{abstract}

In this work, we explore new solutions with static and spherical symmetry in $4D$ for black holes and regular black holes in the recently proposed Conformal Killing Gravity (CKG). This theory is of third order in the derivatives of the metric tensor and essentially satisfies three theoretical criteria for gravitational theories beyond General Relativity (GR). The criteria essentially stipulate the following, that one should: (i) obtain the cosmological constant as an integration constant; (ii) derive the energy conservation law as a consequence of the field equations, rather than assuming it; (iii) and not necessarily consider conformally flat metrics as vacuum solutions.  In fact, existing modified theories of gravity, including GR, do not simultaneously fulfil all of these three criteria.  Here, we couple CKG to nonlinear electrodynamics (NLED) and scalar fields, and we explore solutions of black holes and regular black holes.  More specifically, by solving the field equations of CKG, we find specific forms for the NLED Lagrangian, the scalar field and the field potential, and analyse the regularity of the solutions through the Kretschmann scalar. We find generalizations of the Schwarschild--Reissner-Nordstr\"{o}m--AdS solutions, and consequently further extend the class of (regular) black hole solutions found in the literature.
\end{abstract}
\maketitle
\def\HMS{{\scriptscriptstyle{\rm HMS}}}


\section{Introduction}\label{sec1}

The discovery of the accelerated expansion of the Universe 
\cite{SupernovaSearchTeam:1998fmf,SupernovaCosmologyProject:1998vns} has spurred much research in modified theories of 
gravity as a possible cause of this late-time cosmic speed-up 
\cite{Capozziello:2002rd,Carroll:2003wy}. In fact, a plethora of modified theories of gravity \cite{Nojiri:2006ri,Nojiri:2010wj,Capozziello:2011et,Clifton:2011jh,Capozziello:2011et,CANTATA:2021ktz,Avelino:2016lpj,Harko:2011kv}, with different approaches and applications \cite{Olmo:2011uz,Harko:2011nh,Harko:2018ayt,BeltranJimenez:2017tkd,BeltranJimenez:2019esp,BeltranJimenez:2018vdo,Cai:2015emx,Geng:2011aj}, have been proposed in the literature.
Recently, an interesting theory, that has received some attention, denoted ``Conformal Killing Gravity'' (CKG) has been proposed that satisfies three theoretical criteria for gravitational theories beyond General Relativity (GR) \cite{Harada}. These criteria are essentially the following: (i) obtain the cosmological constant as an integration constant; (ii) derive the energy conservation law as a consequence of the field equations, rather than assuming it; (iii) and not necessarily consider conformally flat metrics as vacuum solutions. It was argued in \cite{Harada} that the existing theories, including GR, do not simultaneously fulfil all of these three criteria. 

Thus, to address this issue, a new gravitational field equation, within CKG (which we will briefly outline below), was proposed that satisfies these points.  The theory was further applied to cosmology, where solutions were obtained that describe the transition from a decelerating to an accelerating expansion in a matter-dominated universe, and thus providing a novel explanation for the current accelerating expansion of the universe \cite{Harada}. It was further shown that the gravitational field equation proposed by Harada in \cite{Harada} is equivalent to an extended Einstein field equation with the presence of an arbitrary conformal Killing tensor \cite{Mantica:2023stl}. This turns Harada's equations of third order in the derivatives of the metric tensor to second order, and offers a simple strategy of obtaining solutions. In this work, we explore solutions of black holes and regular black holes in CKG coupled to scalar fields and nonlinear electrodynamics (NLED).

Relative to the regularity issue, regular black hole solutions were found in the context of NLED \cite{Ansoldi:2008jw,Bardeen,Ayon-Beato:2000mjt}, where the Bardeen model \cite{Bardeen} was the first regular solution encountered in GR. The Bardeen solution was later reinterpreted as the gravitational field of a nonlinear magnetic monopole, namely, it was considered as a magnetic solution to the Einstein field equations coupled to a NLED \cite{Ayon-Beato:2000mjt}. This has motivated much research in regular black holes, and a plethora of methods and procedures have been explored to construct exact black hole solutions with electric or magnetic charges in GR coupled to NLED \cite{Ayon-Beato:2004ywd,Hollenstein:2008hp,Balart:2014jia,Lemos:2011dq,Balart:2014cga,Fan:2016hvf,Junior:2023qaq}. However, it was stressed in \cite{Bronnikov:2017sgg}, that considering spherically symmetric configurations in GR, supported by nonlinear electromagnetic fields with gauge-invariant Lagrangians depending on the single invariant $F=\frac{1}{4}F^{\mu\nu}F_{\mu\nu}$, only pure magnetic solutions can be completely nonsingular. It is also shown that configurations described only by electric charge can be completely nonsingular if the weak field limit of Maxwell's equations is not reached~\cite{Bronnikov:2000vy}.

In this work, we build on the latter work, and explore black hole and regular black hole solutions in CKG coupled to NLED and scalar fields. This work in outlined in the following manner: In Sec. \ref{sec2}, we briefly present the field equations of Conformal Killing Gravity, coupled to NLED and scalar fields, and consider solutions solely described by a magnetic charge.  In Sec. \ref{sec3}, we explore black hole solutions, for specific cases of the scalar field and the scalar potential,  and in Sec. \ref{sec4}, extend the analysis to regular black hole spacetimes. Finally, in Sec.  \ref{sec:concl}, we summarize and discuss our results.


\section{Conformal Killing Gravity coupled to non-linear electrodynamics and scalar fields}\label{sec2}

Recently, a theory denoted by ``Conformal Killing Gravity'' (CKG) has been proposed ~\cite{Harada, Mantica:2023stl}, that generalizes General Relativity (GR).  The gravitational field equations are given by
\begin{equation}
H_{\alpha\mu\nu}=\kappa^2\Theta_{\alpha\mu\nu},\label{eq_CKG}
\end{equation}
where $\kappa^2=8\pi G/c^4$, $G$ is the gravitational constant, and throughout this work we use the natural units, i.e. $G=c=1$. Furthermore, the tensors, $H_{\alpha\mu\nu}$ and $\Theta_{\alpha\mu\nu}$, are defined by
\begin{eqnarray}
H_{\alpha\mu\nu}&\equiv & \nabla_{\alpha}R_{\mu\nu}+\nabla_{\mu}R_{\nu\alpha}+\nabla_{\nu}R_{\alpha\mu}
	\nonumber \\
&&-\frac{1}{3}\left(g_{\mu\nu}\partial_{\alpha}+g_{\nu\alpha}\partial_{\mu}+g_{\alpha\mu}\partial_{\nu}\right)R,
	 \\
\Theta_{\alpha\mu\nu}&\equiv & \nabla_{\alpha}\Theta_{\mu\nu}+\nabla_{\mu}\Theta_{\nu\alpha}+\nabla_{\nu}\Theta_{\alpha\mu}
	\nonumber \\
&&-\frac{1}{6}\left(g_{\mu\nu}\partial_{\alpha}+g_{\nu\alpha}\partial_{\mu}+g_{\alpha\mu}\partial_{\nu}\right)\Theta \,,
\end{eqnarray}
respectively.
Here $R_{\mu\nu}$ stands for the Ricci tensor and $\Theta_{\mu\nu}$  for the energy-momentum tensor, where their traces are represented by $R$ and $\Theta$, respectively.
Moreover, Einstein's equations of GR are solutions of Eq.~\eqref{eq_CKG}, and therefore, GR is a particular case of CKG. Note that if we contract the equation of motion ~\eqref{eq_CKG} by $g^{\mu\nu}$, the conservation law $\nabla_\alpha\Theta^\alpha_{\phantom{\alpha}\mu}=0$ is naturally  satisfied.

We will now explore the contributions of NLED and a scalar field as matter sources applied to the energy-momentum tensor in the field equations~\eqref{eq_CKG}. Our goal is to investigate the description of the metric solutions for black holes and regular black holes, using these forms of matter in the theory of CKG. Considering the above, we can express the energy-momentum tensor by the following description:
\begin{equation}
    \Theta_{\alpha\mu\nu}=\overset{F}{\Theta}_{\alpha\mu\nu}+\overset{\varphi}{\Theta}_{\alpha\mu\nu}.\label{TEM}
\end{equation}
The explicit forms of the components resulting from the contribution of the energy-momentum tensor described by the nonlinear electromagnetic field and the scalar field are defined as follows:
\begin{eqnarray}
\overset{F}{\Theta}_{\alpha\mu\nu}&\equiv & \nabla_{\alpha}\overset{F}{T}_{\mu\nu}+\nabla_{\mu}\overset{F}{T}_{\nu\alpha}+\nabla_{\nu}\overset{F}{T}_{\alpha\mu}
	\nonumber \\
&&-\frac{1}{6}\left(g_{\mu\nu}\partial_{\alpha}+g_{\nu\alpha}\partial_{\mu}+g_{\alpha\mu}\partial_{\nu}\right)\overset{F}{T},\\
\overset{\varphi}{\Theta}_{\alpha\mu\nu}&\equiv &\nabla_{\alpha}\overset{\varphi}{T}_{\mu\nu}+\nabla_{\mu}\overset{\varphi}{T}_{\nu\alpha}+\nabla_{\nu}\overset{\varphi}{T}_{\alpha\mu}
	\nonumber \\
&&-\frac{1}{6}\left(g_{\mu\nu}\partial_{\alpha}+g_{\nu\alpha}\partial_{\mu}+g_{\alpha\mu}\partial_{\nu}\right)\overset{\varphi}{T} \,,
\end{eqnarray}
respectively, with
\begin{align}
    \overset{F}{T}{}_{\mu\nu}&=g_{\mu\nu}{\cal L}_{\rm NED}(F)-{\cal L}_{F}F_{\mu\alpha}F_{\nu}^{\phantom{\nu}\alpha},\\
\overset{F}{T}&=4{\cal L}_{\rm NED}(F)-4{\cal L}_{F}F,\\
    \overset{\varphi}{T}{}_{\mu\nu}&=2\,\epsilon\,\partial_{\mu}\varphi\partial_{\nu}\varphi-\epsilon 
g_{\mu\nu}\partial^{\sigma}\varphi\partial_{\sigma}\varphi+g_{\mu\nu}V(\varphi),\\
\overset{\varphi}{T}&=2\,\epsilon\,\partial^{\nu}\varphi\partial_{\nu}\varphi-\epsilon4\partial^{\sigma}\varphi\partial_{\sigma}\varphi+4V(\varphi).
\end{align}
\par
Here $\epsilon=+1$ is the scalar field, while $\epsilon=-1$ represents the phantom scalar field, $\varphi$ represents the scalar field, while $V(\varphi)$ denotes the scalar potential,  ${\cal L}_{\rm NED}(F)$ is the Lagrangian density describing a NLED that depends of the electromagnetic scalar
\begin{equation}
F=\frac{1}{4}F^{\mu\nu}F_{\mu\nu},    \label{F}
\end{equation}
 and the  Maxwell-Faraday antisymmetric tensor is defined by
\begin{equation}
   F_{\mu \nu} = \partial_\mu A_\nu -\partial_\nu A_\mu, 
\end{equation}
 where ${A_\alpha}$ is the magnetic vector potential. 

Some important expressions resulting from the influence of the gravitational field are as follows:
\begin{align}
\nabla_\mu ({\cal L}_F F^{\mu\nu})&=\sqrt{-g} \partial_\mu (\sqrt{-g} {\cal L}_F F^{\mu\nu})=0,\label{sol2}\\
     2\nabla_\nu\nabla^\mu \varphi&=-\frac{dV(\varphi)}{d\varphi}\,,
\end{align}
where we denote ${\cal L}_F=\partial {\cal L} _{\rm NED}(F)/\partial F$.
\par
Throughout this work we consider the following static and spherically symmetric metric:
\begin{equation}
ds^2=e^{a(r)}dt^2-e^{b(r)}dr^2-\Sigma^2(r) \, d\Omega^2,\label{m}
\end{equation}
where $a(r), b(r)$ and $\Sigma(r)$ are functions of the radial coordinate $r$, and are independent of time, and 
\begin{equation}
d\Omega^2\equiv d\theta^{2}+\sin^{2}\left(\theta\right)d\phi^{2}.
\end{equation}
Note that the metric function $\Sigma(r)$ makes it possible to describe the metric in its most general form for static and spherically symmetric~\eqref{m}. In the sections below, we will provide the corresponding metric for the solutions of black holes and regular black holes.

Furthermore, we consider solutions described solely by the magnetic charge, $q$. The components for $F_{\mu\nu}$ and the electromagnetic scalar are:
\begin{equation}
    F_{23}=q \sin\theta \,,
\end{equation}
and the electromagnetic scalar is given by
\begin{equation}
    F=\frac{q^2}{2 \Sigma^4(r)}.
\end{equation}
\par
To assess the consistency of our solutions, we will also use the following useful relationship
\begin{equation}
    {\cal L}_F-\frac{\partial {\cal L} _{\rm NED}}{\partial r} \bigg(\frac{\partial F}{\partial r}\bigg)^{-1}=0.\label{RC}
\end{equation}

Outlining the strategy that we follow throughout this paper, in finding specific (regular) black hole solutions in CKG is in order. The field equations~\eqref{eq_CKG} are coupled gravitational equations of third order in the metric, so that the inclusion of the material content makes it very difficult to obtain a solution by direct integration. To overcome this difficulty, and to find new solutions for these equations, we will use the following strategy in our work: We propose a symmetry and a functional form for the metric that already includes the new characteristic term of the vacuum solution, $-\lambda r^4/5$, then we integrate the equations of motion to determine the NLED Lagrangian and its derivative, and finally we check the consistency equation~\eqref{RC}. In this way, we generalize several known GR solutions and obtain the material content of this solution in CKG, consequently extending the class of (regular) black hole solutions found in the literature.

\section{Black hole solutions}\label{sec3}

In this section, we study black hole solutions where the matter content in the equations of motion~\eqref{eq_CKG} are supported by NLED and a scalar field. 
However, for simplicity,  in the first two topics discussed in sections~\eqref{Maxwell} and~\eqref{sol_BH2}, we will only present solutions coupled to NLED.

\subsection{Black holes coupled to NLED I: Maxwell type}\label{Maxwell}

In the first two models we consider $\Sigma(r)=r$, so that the metric \eqref{m},  for these models reduces to
\begin{equation}
ds^2=e^{a(r)}dt^2-e^{b(r)}dr^2-r^2\, d\Omega^2.
\label{m_BH}
\end{equation}
In this model we will assume that~$V(\varphi)=0$ and $\varphi=0$, and consider a symmetry where
\begin{equation}
    a(r)=-b(r).
\end{equation}
Furthermore, based on these specifications, we propose the following metric function
\begin{equation}
  e^{a(r)}=e^{-b(r)}=1-\frac{2 M}{r}+\frac{k_0}{r^2}-\frac{\Lambda  r^2}{3}-\frac{\lambda  r^4}{5}.\label{a_Maxwell}
\end{equation}

Thus, after substituting the metric \eqref{m_BH} and the function \eqref{a_Maxwell} to obtain the equations of motion resulting from \eqref{eq_CKG}, we can solve the ``0, 0, 1" and ``2, 1, 2" components of the field equations to obtain the electromagnetic Lagrangian and its derivative, as follows:
\begin{align}
 {\cal L}_{\rm NED}(r) & =-f_1 q^2 r^2+f_0+\frac{k_0}{\kappa ^2 r^4},
 \label{L_Maxwell}\\
 {\cal L}_F (r) &=f_1 r^6+\frac{2 k_0}{\kappa ^2 q^2}.
 \label{LF_Maxwell}
\end{align}
Note that this Lagrangian differs from the Maxwell Lagrangian by the term $r^2$.

From condition~\eqref{RC}, we obtain $f_1 r^6=0$, thus for consistency we impose $ f_1=0$.
Therefore, the expressions for the Lagrangian and its derivative are identically satisfied by the consistency relation~\eqref{RC} and also satisfied by the equations of motion. This implies that we have the linear Maxwell case.

Let us define the constant in Eq.~\eqref{a_Maxwell} as $k_0=q^2$, i.e., as magnetic charge, so that we obtain
\begin{equation}
    e^{a(r)}=e^{-b(r)}=1-\frac{2 M}{r}+\frac{q^2}{r^2}-\frac{\Lambda  r^2}{3}-\frac{\lambda  r^4}{5}.\label{a2_Maxwell}
\end{equation}
Note that this metric function contains an additional term with $\lambda$ which classifies it as a Maxwell-Reissner-Nordstr\"{o}m-AdS type solution. Consequently, we obtain the Maxwell-Reissner-Nordstr\"{o}m-AdS solution by taking into account $\lambda\rightarrow0$.

From Eq.~\eqref{F}, we can obtain $r(F)$, and in this way it is possible to express the following
\begin{align}
    {\cal L}(F)&=\frac{2 F}{\kappa ^2},\label{L3_BN}\\
    {\cal L}_F(F)&=\frac{2 }{\kappa ^2},
\end{align}
which describes the material content that supports the black hole geometry.
\par
The Kretschmann scalar is given by
\begin{eqnarray}
    K &=& \frac{8 \Lambda ^2}{3}+\frac{48 M^2}{r^6}-\frac{8 q^2 \left(60 M+11 \lambda  r^5\right)}{5 r^7}
	\nonumber  \\    
    &&+\frac{48 \lambda  M}{5 r}+\frac{56 q^4}{r^8}+\frac{212 \lambda ^2 r^4}{25}+8 \lambda  \Lambda  r^2,
\end{eqnarray}
which diverges for $r\rightarrow0$ and $r\rightarrow\infty$,  and considering $\lambda = 0$, only $r\rightarrow0$ remains as a divergence.

The curvature scalar is given by 
\begin{equation}
    R=-4 \Lambda -6 \lambda  r^2.
\end{equation}
where the divergence disappears if $\lambda\rightarrow0$.

Note that the field equation \eqref{eq_CKG} generalizes the Einstein field equation, so for a particular case, it should reduce to GR.  More specifically, the trace of \eqref{eq_CKG} for this Reissiner-Nordstr\"{o}m-type solution takes the form $\kappa^2 \Theta+R=-4 \Lambda -6 \lambda  r^2$, which reduces to the trace of the Einstein field equation, given  by $\kappa^2 T+R=-4\Lambda$,  when $\lambda=0$.

\subsection{Black holes coupled to NLED II}\label{sol_BH2}

In this second approach, as mentioned above we only consider solutions with NLED, i.e., $V(\varphi)=0$ and $\varphi=0$. Thus, we consider again the metric \eqref{m_BH} to obtain the equations of motion of \eqref{eq_CKG}, and taking into account the components ``0, 0, 1'' and ``2, 1, 2'' of the resulting equations of motion, we obtain:
\begin{widetext}
\begin{eqnarray}
 {\cal L}_{\rm NED} (r) &=& -f_{0}q^{2}r^{2}+f_{1}-q^{2}r^{2}\int\Bigg\{ -\frac{4-4e^{-b(r)}}{\kappa^{2}q^{2}r^{5}}-\frac{e^{-b(r)}a'(r)}{4\kappa^{2}q^{2}r^{4}}\Big[r^{2}\Big(b''(r)-2a''(r)\Big)-r^{2}b'^{2}(r)+r4b'(r)-2\Big]
 	\nonumber\\
 && -\frac{re^{-b(r)}}{4\kappa^{2}q^{2}r^{5}}\left[3\left(r^{2}a''(r)-2\right)b'(r)+ra'^{2}(r)\left(rb'(r)+6\right)-2r\left(ra^{(3)}(r)-a''(r)+b''(r)\right)+2rb'^{2}(r)\right]\Bigg\} dr 
 	\nonumber\\
 && +\int\Bigg\{ \frac{e^{-b(r)}}{2\kappa^{2}r^{2}}\left[-6ra'^{2}(r)+r\left(ra^{(3)}(r)+2a''(r)+4b''(r)-4b'^{2}(r)\right)+a'(r)\left(r^{2}a''(r)-2\right)\right]
 	\nonumber\\
&& +\frac{4e^{-b(r)}-4}{\kappa^{2}r^{3}}+\frac{e^{-b(r)}}{4\kappa^{2}r}\Bigl[b'(r)(a'(r)\Bigl(-ra'(r)+rb'(r)-16\Bigr)-3ra''(r)\Bigr)-ra'(r)b''(r)\Bigr]\Bigg\} dr\,,	
\label{L_BH2}
\end{eqnarray}
\begin{eqnarray}
 {\cal L}_F (r) &=&  f_{0}r^6+r^{6}\int\Bigg\{ -\frac{4-4e^{-b(r)}}{\kappa^{2}q^{2}r^{5}}-\frac{e^{-b(r)}a'(r)}{4\kappa^{2}q^{2}r^{4}}\left[r^{2}\left(b''(r)-2a''(r)\right)-r^{2}b'^{2}(r)+r4b'(r)-2\right]
 	\nonumber\\
 && \hspace{-1cm} -\frac{re^{-b(r)}}{4\kappa^{2}q^{2}r^{5}}\left[3\left(r^{2}a''(r)-2\right)b'(r)+ra'^{2}(r)\left(rb'(r)+6\right)-2r\left(ra^{(3)}(r)-a''(r)+b''(r)\right)+2rb'^{2}(r)\right]\Bigg\} dr.\label{LF_BH2}
\end{eqnarray}
\end{widetext} 
 
Equations \eqref{L_BH2} and \eqref{LF_BH2} do not satisfy all of the equations of motion, and thus taking into account the component ``1,1,1" of the field equation, we obtain:
%
\begin{eqnarray}
-r \left[a''(r)+b''(r)\right]+a'(r) \left[2 r b'(r)+1\right]
	\nonumber \\
+r a'(r)^2+r b'(r)^2+b'(r)=0 \,.
	\label{111}
\end{eqnarray}

As a working hypothesis, consider the following functional form for $b(r)$
\begin{equation}
    b(r)=	\begin{cases}
-a(r),\\
-a(r)-\ln\left(b_{0}+r^{2}\right)\,,
\end{cases}
    \label{b_BH2}
\end{equation}
where $b_0$ is a constant. 
The symmetry $b(r)=	
-a(r)$ is well known and is used in numerous solutions for black holes in general relativity. However, it is also possible to find solutions in which this symmetry is relaxed, as observed for example in~\cite{Junior:2023qaq,Junior:2015dga,Junior:2015fya}.

We analyse both conditions in detail below.

\subsubsection{Model with $b(r)=-a(r)$}

For this specific case,  all equations of motion are fulfilled, as well as the condition \eqref{RC}.
To develop our solutions further, we assume that the metric function $a(r)$ is described by
\begin{equation}
     e^{a(r)}=e^{-b(r)}= 1-\frac{2M}{r}+\frac{q^2}{r^2}-\frac{\Lambda r^{2}}{3}-\frac{\lambda r^{4}}{5}+\frac{k_{0}}{r^{4}} .\label{a_BH2}
\end{equation}
This metric function is of the Reissner-Nordstr\"{o}m-AdS type, as it contains an additional term with $\lambda$ and $k_0$. 

Since we are dealing with a static and spherically symmetric metric in which $g_{tt}=-g_{rr}^{-1}$ holds according to the metric~\eqref{m}, the condition for the formation of an event horizon is as follows:
\begin{equation}
     e^{a(r_{H})}=0,\label{rH}
\end{equation}
where $r_{H}$ is the radius of the horizon.  It is extremely difficult to obtain analytic solution of Eq. (\ref{rH}) for the horizon radius algebraically. However, we can address the problem numerically by assigning specific values to the parameters so that we can explore the presence of event horizons through graphical representations. 
In this context, in order to classify the horizons, we consider the following  condition:
\begin{equation}
    \frac{d e^{a(r)}}{dr}\bigg|_{r=r_H}=0.\label{der_a}
\end{equation}
which provides relevant information on the critical values of the model parameters, namely, $M$, $q$, $\Lambda$, $\lambda$ or $k_0$.
Thus, Eqs. \eqref{rH} and \eqref{der_a} provide algebraic solutions for the radius of the horizon and the critical parameters. However, these solutions are complicated and difficult to handle, so we will solve these expressions numerically by assigning specific values to the parameters. 
To gain some insight into the solutions and for simplicity, we will focus on AdS-type solutions, i.e., $\Lambda < 0$, and with $\lambda= 0$. In fact, as will be shown below, this choice ensures that the Kretschmann scalar \eqref{K_BH2} is nonsingular.

\begin{figure}[!h]
\centering
\includegraphics[scale=0.41]{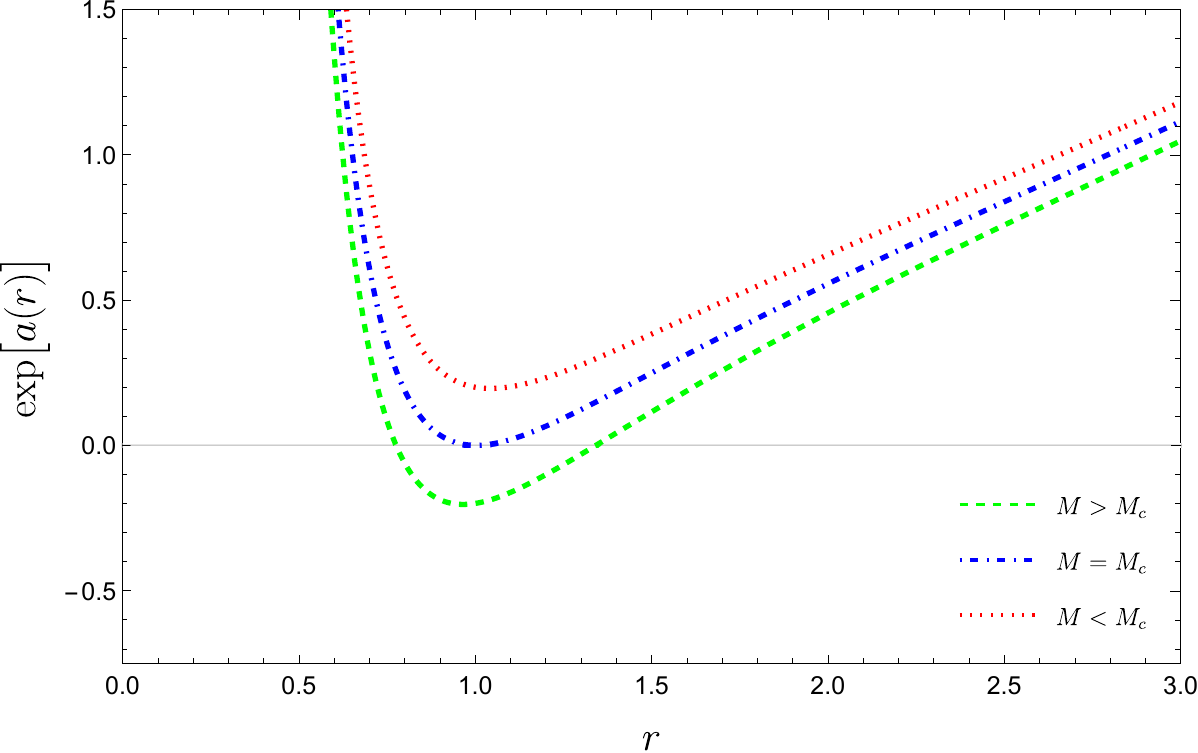}
\caption{The plot depicts the metric function \eqref{a_BH2}, i.e., $\exp{[a(r)]}$, for the specific values of $\lambda=0$, $k_0=0.4$, $\Lambda =-0.2$ and $q=0.1$, where Eq. (\ref{der_a}) determines the value of the critical mass, given by $M_c = 0.738$.  For $M > M_c$ there are two horizons, for $M = M_c$, there is only one horizon and for $M < M_c$, there is no horizon. } 
\label{figM_BH2}
\end{figure}

More specifically, to obtain numerical solutions, and as a first example, we set the following values for the parameters: $q = 0.1$, $\Lambda = -0.2$, and $k_0 = 0.4$. In this way, we simultaneously establish the two conditions~\eqref{rH} and~\eqref{der_a} for the case of an extreme solution, by obtaining the values of $r_H$ and the critical mass $M_c$. In this particular scenario, we obtain $M_c = 0.738$, and the number of horizons depends directly on the values of these parameters. 
Figure~\ref{figM_BH2} depicts the behavior of the metric function~\eqref{a_BH2}, where the mass takes the values $M > M_c$, $M = M_c$, and $M < M_c$. 
For $M > M_c$ there are two horizons, where the innermost one is the Cauchy horizon and the second is the event horizon; for $M = M_c$, there is only one horizon; and for $M < M_c$, one verifies the absence of event horizons. 

\begin{figure}[!h]
\centering
\includegraphics[scale=0.41]{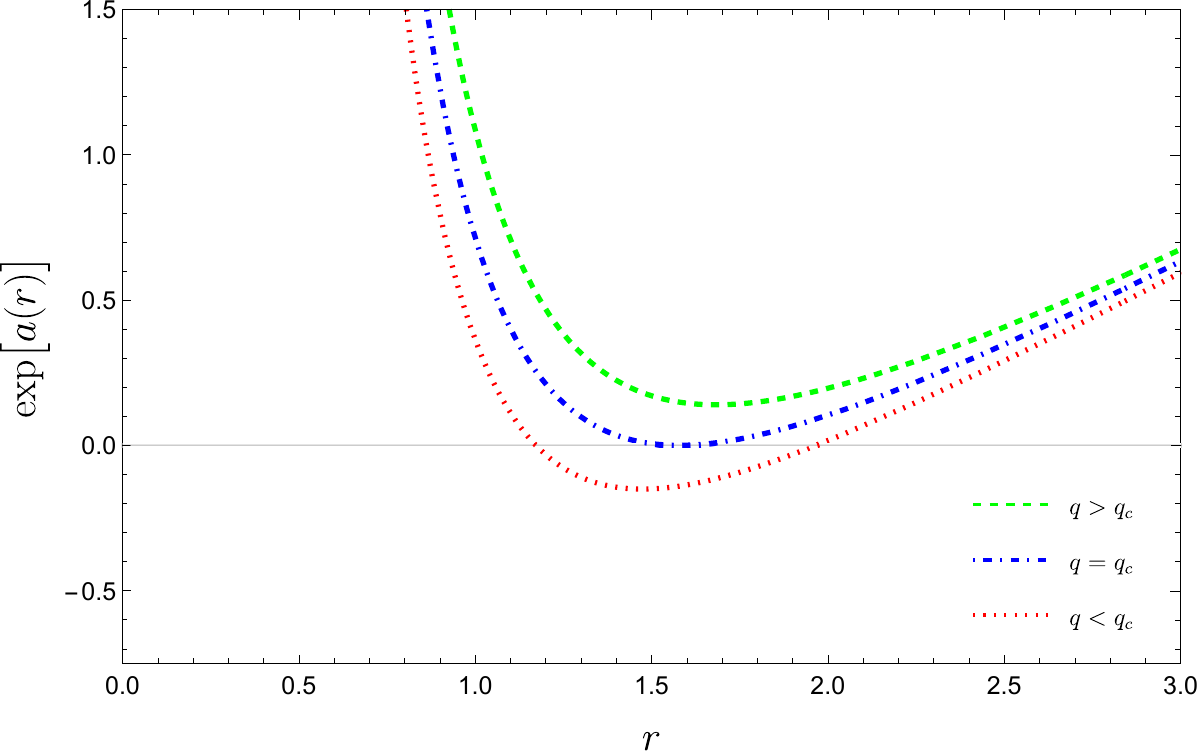}
\caption{The plot depicts $\exp{[a(r)]}$, given by Eq.~\eqref{a_BH2}. We consider $\lambda=0$, $M=2$, $\Lambda =-0.2$ and $k_0=0.4$, where the critical value of the charge is given by $q_c=1.802$. For $q > q_c$, there are no horizons; if $q=q_c$, there is only one horizon; and for $q<q_c$ two horizons exist, where the innermost is the Cauchy horizon and the outermost, the event horizon, respectively.} 
\label{figq_BH2}
\end{figure}

For the second numerical solution, we now set the following values for the constants: $M=2.0$, $\Lambda=-0.2$, and $k_0=0.4$. This leads us to the critical charge value of $q_c=1.802$. The behavior of $\exp{[a(r)]}$ as a function of $r$ is depicted in Fig.~\ref{figq_BH2}, for the cases $q > q_c$, $q=q_c$, and $q<q_c$. Here,when the charge is greater than the critical charge, there are no horizons; when the charge is equal to the critical charge, there is only one horizon; and if the charge is less than the critical charge, there are two horizons: the innermost Cauchy horizon and the outermost event horizon.

\begin{figure}[!h]
\centering
\includegraphics[scale=0.41]{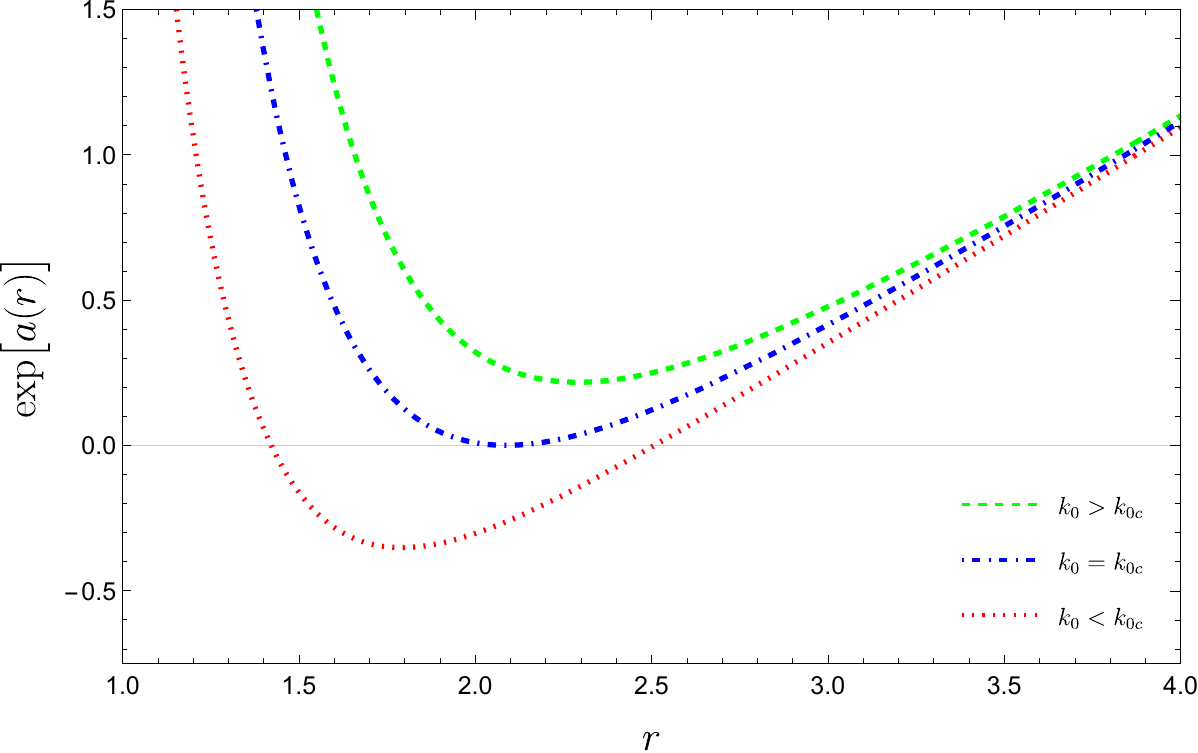}
\caption{The plot depicts the expression $\exp{[a(r)]}$, given by Eq.~\eqref{a_BH2}. We consider $\lambda=0$, $M=2$, $\Lambda =-0.2$ and $q=0.1$, so that $k_{0c}=11.837$.  Note the absence of horizons for $k_0 > k_{c0}$; for $k_0=k_{c0}$, there is only one horizon; and if $k_0 < k_{c0}$, two horizons are present.} 
\label{figk0_BH2}
\end{figure}

Finally, for the following values of the parameters: $M=2.0$, $\Lambda=-0.2$ and $q=0.1$, we obtain $k_{0c}=11.837$. For the values of $k_0 > k_{0c}$, $k_0=k_{0c}$ and $k_0<k_{0c}$, the behavior of $\exp{[a(r)]}$ as a function of $r$ is shown in Fig.~\ref{figk0_BH2}.   We verify that there are no horizons for $k_0 > k_{c0}$; if $k_0=k_{c0}$, there is only one horizon; and if $k_0 < k_{c0}$, there are two horizons.
\par

Now, to determine the NLED Lagrangian, by substituting the metric function \eqref{a_BH2} into Eqs. \eqref{L_BH2} and \eqref{LF_BH2}, we find
\begin{align}
 {\cal L}_{\rm NED} (r) &= f_0-f_1 q^2 r^2+\frac{q^2}{\kappa ^2 r^4}+\frac{3 k_0}{\kappa ^2 r^6}\,,
 	\label{L2_BH2}\\
 {\cal L}_F (r) &=\frac{2}{\kappa ^2}+ r^6 \left(f_1+\frac{9 k_0}{\kappa ^2 q^2 r^8}\right).
 	\label{LF2_BH2}
\end{align}
It is straightforward to show that Eqs.~\eqref{L2_BH2} and~\eqref{LF2_BH2} satisfy the relation~\eqref{RC} and all the equations of motion.

The Kretschmann scalar is given by
\begin{eqnarray}
    K &=& \frac{8 \Lambda ^2}{3}+\frac{468 k_0^2}{r^{12}}-\frac{8 k_0 \left(30 M r-39 q^2+9 \lambda  r^6+\Lambda  r^4\right)}{r^{10}}
	\nonumber \\    
   && +\frac{8 \left(6 M^2 r^2-12 M q^2 r+7 q^4\right)}{r^8}+\frac{8 \lambda  \left(6 M r-11 q^2\right)}{5 r^2}
    \nonumber\\
    &&+\frac{212 \lambda ^2 r^4}{25}+8 \lambda  \Lambda  r^2\,,
    \label{K_BH2}
\end{eqnarray}
which diverges in the limit~$r\rightarrow0$ and ~$r\rightarrow\infty$. If we consider $\lambda=0$, only the divergence of the limit of $r\rightarrow0$ remains.

The trace of the field equations yields
\begin{equation}
    \kappa^2\Theta+R=4 f_0 \kappa ^2-6 f_1 \kappa ^2 q^2 r^2-4 \Lambda -6 \lambda  r^2\,,
    \label{scalars_BH2}
\end{equation}
which reduces to GR for  $f_0=f_1=\lambda=0$.

\begin{figure}[!h]
\centering
\includegraphics[scale=0.45]{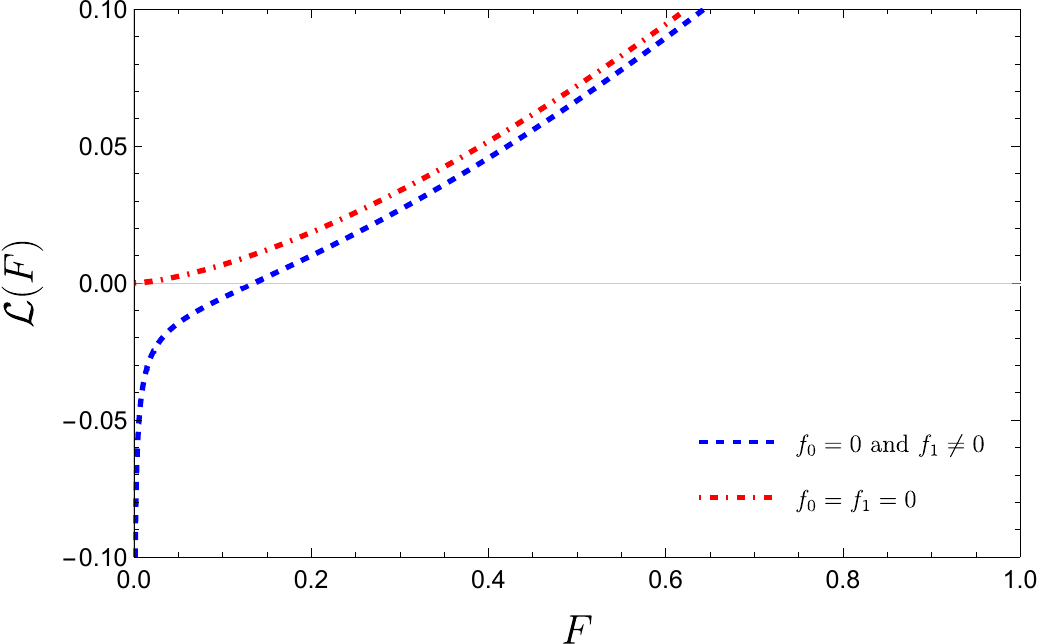}
\caption{Graphical representation of the expression ${\cal L}(F) $, describe by Eq.~\eqref{L3_BH2}. The blue dashed curve represents the behavior of ${\cal L}(F)$ versus $F$ with $f_0=0$ and $f_1=0.2$, while the red dot-dashed curve illustrates the behavior of ${\cal L}(F)$ versus $F$ with $f_0=f_1=0$. The values of the parameters used are $k_0=0.4$, $q=0.3$ and $\kappa=8\pi$. } 
\label{LxF_BH2}
\end{figure}

Taking into account Eq. ~\eqref{F}, we can determine the inverse function  $r(F)$, so that the electromagnetic Lagrangian is given by
\begin{equation}
    {\cal L}(F)=f_0-\frac{f_1 q^3}{ \sqrt{2 F}}+\frac{2 F \left(3  \sqrt{2 F} k_0+q^3\right)}{\kappa ^2 q^3}\,,\label{L3_BH2}
\end{equation}
which satisfies all the equations of motion and also the relation~\eqref{RC}.  Note that we recover the Reissner-Nordstr\"{o}m Lagrangean for the specific case of $f_0=f_1=k_0=0$, as expected.
The behavior of Eq.~\eqref{L3_BH2} is depicted in Fig.~\eqref{LxF_BH2}. The blue dashed curve represents the values of $f_0=0$ and $f_1=0.2$, while the dotted-dashed curve shows the behavior for the values of $f_0=0$ and $f_1=0$, which corresponds to the behavior described by GR.

We note that this Lagrangian,  Eq.~\eqref{L3_BH2}, grows indefinitely when we consider the limit $F\gg1$. In this scenario, the Lagrangian ${\cal L}(F)$ of the CKG theory approaches GR, as can be seen from the curves in Fig.~\ref{LxF_BH2}, so 
\begin{equation}
{\cal L}(F)\sim f_{0}+\frac{6F\sqrt{2F}k_{0}}{\kappa^{2}q^{3}}.
\end{equation}
If we consider the limit $F\ll1$, we notice a difference between the theories, as the GR curve tends to zero, while the curve described by CKG leads to a divergent Lagrangian in this limit
\begin{equation}
{\cal L}(F)\sim f_{0}-\frac{f_{1}q^{3}}{\sqrt{2F}}.
\end{equation}

\par

\subsubsection{Model with $b(r)=-a(r)-\ln\left(b_{0}+r^{2}\right)$}

For this specific we will use the metric function $a(r)$ \eqref{a_BH2} described in the previous model in~\eqref{b_BH2}, we obtain
\begin{equation}
    e^{a(r)}=\left(1-\frac{2M}{r}+\frac{q^2}{r^2}-\frac{\Lambda r^{2}}{3}-\frac{\lambda r^{4}}{5}+\frac{k_{0}}{r^{4}}\right),
     \label{b1_BN2}
\end{equation}
and
\begin{equation}
    e^{-b(r)}=\left(b_0+r^2\right) \left(1-\frac{2M}{r}+\frac{q^2}{r^2}-\frac{\Lambda r^{2}}{3}-\frac{\lambda r^{4}}{5}+\frac{k_{0}}{r^{4}}\right)\,,
    \label{expb}
\end{equation}
which satisfy all the equations of motion.
Taking into account Eqs.  \eqref{b1_BN2} and \eqref{expb}, the Lagrangian \eqref{L_BH2} and its derivative \eqref{LF_BH2} take the form:
\begin{eqnarray}
 {\cal L}_{\rm NED} (r) &=& f_0-f_1 q^2 r^2+\frac{1}{5 \kappa ^2 r^6}
\Big\{ 5 \Big[r^2 \left(b_0 q^2+4 k_0\right)
	\nonumber \\ 
 && \hspace{-1.75cm} +3 b_0 k_0 +r^4 \left(-b_0+2 q^2+1\right)-2 M r^5\Big]+4 \lambda  r^{10}\Big\}\,,
 \label{L4_BH2}
\end{eqnarray}
\begin{eqnarray}
 {\cal L}_F (r) &=& r^6 \Bigg\{f_1+
	\frac{1}{5 \kappa ^2 q^2 r^8}\Big[ 
 10 r^2 \left(b_0 q^2+4 k_0\right)+45 b_0 k_0
	\nonumber \\ 
&& -5 r^4 \left(b_0-2 q^2-1\right)-5 M r^5-8 \lambda  r^{10} \Big] \Big\}\,,
 \label{LF3_BH2}
\end{eqnarray}
which also satisfy condition~\eqref{RC}. 

In this case, we note that the Kretschmann scalar diverges in the limit of~$r\rightarrow\infty$, and is regular for this limit when we consider that $\lambda=0$. The trace of the field equation is given by
\begin{eqnarray}
    \kappa^2\Theta+R = -4 b_0 \Lambda -6 b_0 \lambda  r^2-\frac{6 q^2}{r^2} \left(f_1 \kappa ^2 r^4-1\right)
	\nonumber \\    
     +4 f_0 \kappa ^2+\frac{6 k_0}{r^4}-\frac{12 M}{r}-\frac{1}{5} 6 \lambda  r^4-6 \Lambda  r^2+6\,,
     \label{scalars2_BH2}
\end{eqnarray}
which does not reduce to GR  for $f_0=f_1=k_0=b_0=\lambda=0$.

\begin{figure}[!h]
\centering
\includegraphics[scale=0.45]{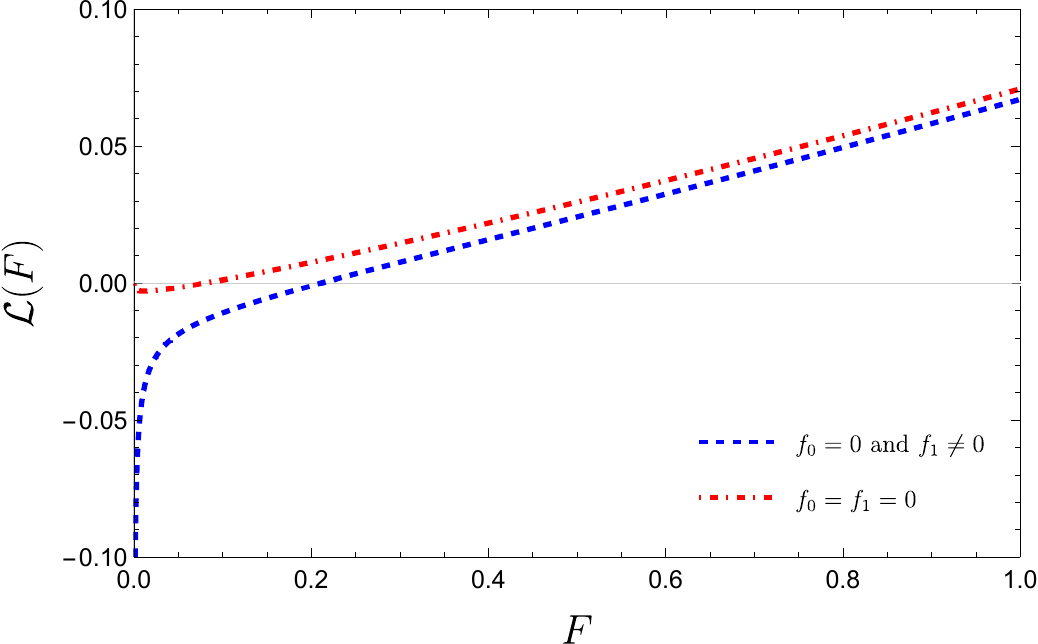}
\caption{The plot represents the function ${\cal L}(F) $, given by Eq.~\eqref{L5_BH2}. The blue dashed curve represents the behavior of ${\cal L}(F)$ versus $F$ with $f_0=0$ and $f_1=0.2$, while the red dotted-dashed curve illustrates the behavior of ${\cal L}(F)$ for $f_0=f_1=0$. The values of the constants used are $\lambda=0$, $k_0=0.4$, $b_0=0.1$,  $q=0.3$, $\kappa=8\pi$ and $M=2.0$. } 
\label{LxF2_BH2}
\end{figure}

The Lagrangean written in terms of $F$ in this model now takes the form 
\begin{eqnarray}
{\cal L}(F)&=& \frac{1}{10F\kappa^{2}q^{2}} \Big[
-5\sqrt{2}f_{1}\sqrt{F}\kappa^{2}q^{5}-20\sqrt[4]{2}F^{5/4}Mq^{3/2}
	\nonumber \\
&&+10\sqrt{2}F^{3/2}q\left(2q^{2}+1\right)+80F^{2}k_{0}+4\lambda q^{4}  \Big]
	\nonumber\\
&& +\frac{b_{0}\sqrt{F}}{\kappa^{2}q^{3}}
\left(6\sqrt{2}Fk_{0}+2\sqrt{F}q^{3}-\sqrt{2}q^{2}\right)
 + f_{0}.\label{L5_BH2}
\end{eqnarray}
Moreover, it satisfies all of the equations of motion and Eq.~\eqref{RC}.
Equation~\eqref{L5_BH2} is depicted in Fig.~\ref{LxF2_BH2}, where the blue dashed curve represents its behavior for the values $f_0=0$ and $f_1=0.2$, and the dotted-dashed curve represents the values of $f_0=0$ and $f_1=0$.

The Lagrangian limit for $F\gg 1$, similar to the previous model, grows indefinitely, and now we have  
\begin{equation}
  {\cal L}(F)  \sim f_{0}+\frac{8k_{0}F}{\kappa^{2}q^{2}}+\frac{6\sqrt{2}k_{0}b_{0}F^{3/2}}{\kappa^{2}q^{3}}\,.
  \label{L6_BH2}
\end{equation}
The only difference now is that in the limit of $F\gg 1$, a very small negative value can be observed in the curve representing the Lagrangian of GR, for instance,  see Fig.~\ref{LxF2_BH2}. The Lagrangian in this limit takes the form
\begin{equation}
{\cal L}(F)\sim f_{0}-\frac{\sqrt{2}f_{1}q^{3}}{2\sqrt{F}}-\frac{\sqrt{2}b_{0}\sqrt{F}}{\kappa^{2}q}.\label{L7_BH2}
\end{equation}
To obtain Eqs.~\eqref{L6_BH2} and~\eqref{L7_BH2}, we consider $\lambda=\Lambda=0$.

\subsection{Third black hole solution}\label{sol_BH3}

In this model, we consider NLED and a scalar field as matter sources, i.e.~$V(\varphi)\neq0$ and $\varphi\neq0$, with the following form for the metric
\begin{equation}
ds^2=e^{a(r)}dt^2-e^{b(r)}dr^2-\bigl(q^2 + r^2\bigr)\, d\Omega^2,
\label{m2_BH}
\end{equation}
where $q$ is assumed to be the magnetic charge. 

Taking into account the equations of motion \eqref{eq_CKG} based on these conditions, we solve the ``0,0,1" and ``2,1,1" components of the resulting equations and thus find the following expressions for ${\cal L}_{\rm NED }$ and ${\cal L}_F$:
\begin{widetext}
\begin{eqnarray}
 &&{\cal L}_{\rm NED} (r) = -q^2 r^2 \int\frac{e^{-b(r)}}{4\kappa^{2}q^{2}\left(q^{2}+r^{2}\right)^{4}}\Bigg\{ 16r\left[r^{2}-e^{b(r)}\left(q^{2}+r^{2}\right)\right]-\left(q^{2}+r^{2}\right)^{2}a'(r)^{2}\left[\left(q^{2}+r^{2}\right)b'(r)+6r\right]\nonumber\\
 &&+a'(r)\left\{ \left(q^{2}+r^{2}\right)^{2}\left[\left(q^{2}+r^{2}\right)\left(2a''(r)-b''(r)+8\kappa^{2}\epsilon\varphi'(r)^{2}\right)+\left(q^{2}+r^{2}\right)b'(r)^{2}-4rb'(r)\right]+2\left(q^{2}+r^{2}\right)\left(5q^{2}+r^{2}\right)\right\}  
 	\nonumber\\
&&+\left(q^{2}+r^{2}\right)^{2}\left[2a^{(3)}(r)\left(q^{2}+r^{2}\right)-2r\left(a''(r)-b''(r)+8\kappa^{2}\epsilon\varphi'(r)^{2}\right)+3b'(r)\left[2-\left(q^{2}+r^{2}\right)a''(r)\right]-2rb'(r)^{2}\right]\Bigg\} dr
	\nonumber\\
 && +\int\left\{ \frac{2\epsilon e^{-b(r)}\varphi'(r)^{2}\left[\left(q^{2}+r^{2}\right)\left(2q^{2}+3r^{2}\right)a'(r)+\left(q^{2}+r^{2}\right)^{2}b'(r)+2q^{2}r\right]}{\left(q^{2}+r^{2}\right)^{2}}-4\epsilon e^{-b(r)}\varphi'(r)\varphi''(r)-V'(r)
 	\right.\nonumber\\
 && -\frac{4re^{-b(r)}\left[q^{2}r^{2}\left(e^{b(r)}-3\right)+r^{4}\left(e^{b(r)}-1\right)-3q^{4}\right]}{\kappa^{2}\left(q^{2}+r^{2}\right)^{4}}+\frac{e^{-b(r)}\left[r\left(3q^{2}+4r^{2}\right)b''(r)-\left(3q^{2}r+4r^{3}\right)b'(r)^{2}+9q^{2}b'(r)\right]}{2\kappa^{2}\left(q^{2}+r^{2}\right)^{2}}
 	\nonumber\\
 && + \frac{re^{-b(r)}\left\{ b'(r)\left[a'(r)\left[-r\left(q^{2}+r^{2}\right)\left(a'(r)-b'(r)\right)-4\left(3q^{2}+4r^{2}\right)\right]-3r\left(q^{2}+r^{2}\right)a''(r)\right]-r\left(q^{2}+r^{2}\right)a'(r)b''(r)\right\} }{4\kappa^{2}\left(q^{2}+r^{2}\right)^{2}}
 	\nonumber\\ 
 && +\frac{r\left(q^{2}+r^{2}\right)\left(ra^{(3)}(r)\left(q^{2}+r^{2}\right)+\left(3q^{2}+2r^{2}\right)a''(r)\right)+a'(r)\left[r^{2}\left(q^{2}+r^{2}\right)^{2}a''(r)+9q^{4}+11q^{2}r^{2}-2r^{4}\right]}{e^{b(r)}2\kappa^{2}\left(q^{2}+r^{2}\right)^{3}}
 	\nonumber\\
 && +\left.\frac{-3r\left(q^{2}+r^{2}\right)\left(q^{2}+2r^{2}\right)a'(r)^{2}}{e^{b(r)}2\kappa^{2}\left(q^{2}+r^{2}\right)^{3}}\right\} dr+f_1-f_0 q^2 r^2 \,, 
 	\label{L_BH3}
 \end{eqnarray}
 \begin{eqnarray}
&&{\cal L}_F (r) = \left(q^2+r^2\right)^3 \int\frac{e^{-b(r)}}{4\kappa^{2}q^{2}\left(q^{2}+r^{2}\right)^{4}}\Bigg\{ 16r\left[r^{2}-e^{b(r)}\left(q^{2}+r^{2}\right)\right]-\left(q^{2}+r^{2}\right)^{2}a'(r)^{2}\left[\left(q^{2}+r^{2}\right)b'(r)+6r\right]
	\nonumber\\
 &&+a'(r)\left\{ \left(q^{2}+r^{2}\right)^{2}\left[\left(q^{2}+r^{2}\right)\left(2a''(r)-b''(r)+8\kappa^{2}\epsilon\varphi'(r)^{2}\right)+\left(q^{2}+r^{2}\right)b'(r)^{2}-4rb'(r)\right]+2\left(q^{2}+r^{2}\right)\left(5q^{2}+r^{2}\right)\right\}  
 	\nonumber\\
&&+\left(q^{2}+r^{2}\right)^{2}\left[2a^{(3)}(r)\left(q^{2}+r^{2}\right)-2r\left(a''(r)-b''(r)+8\kappa^{2}\epsilon\varphi'(r)^{2}\right)+3b'(r)\left[2-\left(q^{2}+r^{2}\right)a''(r)\right]-2rb'(r)^{2}\right]\Bigg\} dr
	\nonumber\\
&& \hspace{1cm}
+\left(q^2+r^2\right)^3f_0 \,.
\label{LF_BH3}
\end{eqnarray}
\end{widetext}
respectively.

To distinguish from the choice $b(r)=-a(r)$, we now describe our metric function as follows
\begin{equation}
    b(r)=-a(r)+b_0(r).\label{relation}
\end{equation}
At this point, we can draw inspiration from the `asymptotically safe gravity-type'~\cite{Bonanno:2000ep}  and describe our metric function in a similar manner, given by:
\begin{equation}
   e^{a(r)}=\left(1-\frac{2 M r}{q^2+r^2}\right).
   \label{a_Bh3}
    \end{equation}
    
Substituting Eqs. \eqref{relation} and \eqref{a_Bh3} into Eqs. \eqref{L_BH3} and \eqref{LF_BH3} and consequently into the equations of motion, we verify that only the component ``1,1,1" is not satisfied. The latter's form is given by:
\begin{eqnarray}
-\frac{3 b_0'(r) \left[r \left(q^2+r^2\right)b_0'(r)-3 q^2+r^2\right]}{\left(q^2+r^2\right)^2}
	\nonumber \\
+\frac{3 r \left[\left(q^2+r^2\right)^2 b_0''(r)+8 q^2\right]}{\left(q^2+r^2\right)^3}
	\nonumber\\
+6 \kappa ^2 \epsilon  \varphi '(r) \Bigl[b_0'(r) \varphi '(r)-2 \varphi ''(r)\Bigr]=0\,.\label{111_BH3}
\end{eqnarray}

\begin{figure}[!h]
\centering
\includegraphics[scale=0.45]{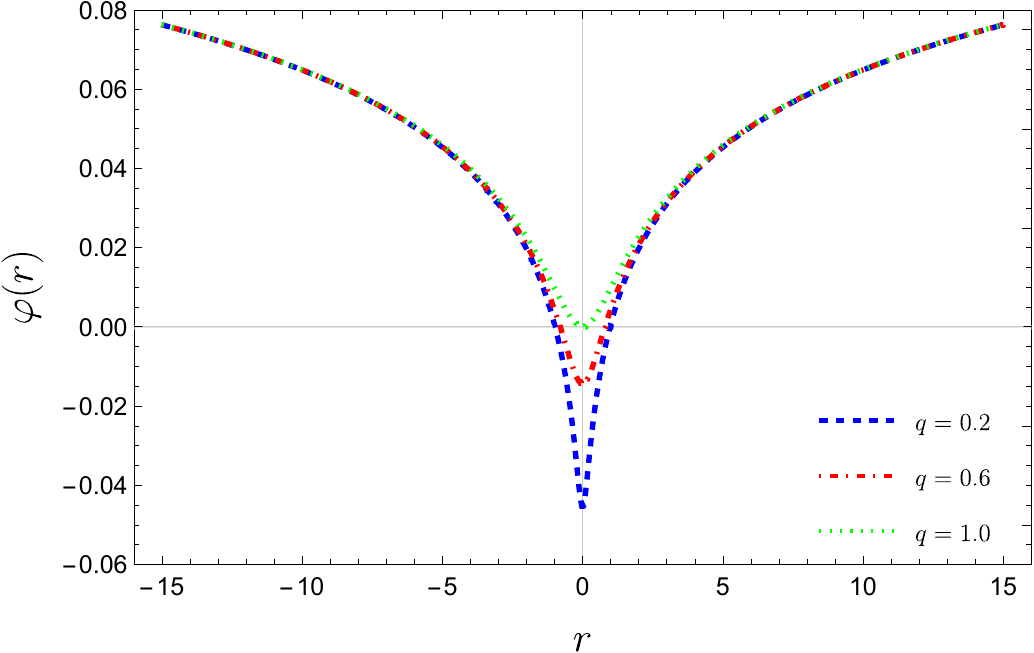}
\caption{Graphical representation of the scalar field $\varphi(r)$, describe by Eq.~\eqref{phi_BH3}, where the charge takes the values of $q=0.2$, $q=0.6$ and $q=1.0$. We also consider $\kappa=8\pi$ and  $\epsilon =1.0$.} 
\label{figphi_BH3}
\end{figure}

For this case we now consider the following functional form
\begin{equation}
    b_0(r)=c_0+2 \log (r)-\frac{1}{2} \log \left(q^2+r^2\right).\label{b0_BH3}
\end{equation}
where $c_0$ is a constant.
The resulting expression after substituting~\eqref{b0_BH3} into~\eqref{111_BH3} can be solved to determine the scalar field, which is given by
\begin{equation}
    \varphi (r) =\frac{\ln \left(q^2+r^2\right)}{2 \kappa\sqrt{2\epsilon }}.\label{phi_BH3}
\end{equation}
In Fig.~\ref{figphi_BH3} we show the behavior of the scalar field~\eqref{phi_BH3} with respect to the radial coordinate $r$ for three different charge values - $q=0.2$, $q=0.6$ and $q=1.0$.
\par
Note that the scalar field consistently takes negative values near the origin of the radial coordinate when the modulus of the charge is in the range 0 to 1, both in the regions $r < 0$ and $r > 0$. For large values of the modulus of the radial coordinate, this scalar field is always positive. 

Therefore, the quantities $V(r)$, ${\cal L} _{\rm NED }(r)$ and ${\cal L} _F(r)$ are now expressed as follows:
\begin{equation}
    V(r)=\frac{-5 q^4 r (12 M+r)-70 M q^2 r^3-28 M r^5-5 q^6}{30 \kappa ^2 q^4 \left(q^2+r^2\right)^{5/2}} \,,
\end{equation}
\begin{eqnarray}    
    {\cal L} _{\rm NED}(r)&=&\left(q^{2}+r^{2}\right)^{3}\left[f_{1}+\frac{M\left(2q^{4}+11q^{2}r^{2}-3r^{4}\right)}{2\kappa^{2}q^{2}r^{3}\left(q^{2}+r^{2}\right)^{7/2}}\right.	
    \nonumber\\
&& \left.+\frac{r^{3}\left(q^{2}+r^{2}\right)\left(2\sqrt{q^{2}+r^{2}}-1\right)}{2\kappa^{2}q^{2}r^{3}\left(q^{2}+r^{2}\right)^{7/2}}\right] \,,
\end{eqnarray}
\begin{eqnarray}
    {\cal L} _{F}(r)&=&f_0 - f_1 q^2 r^2-\frac{3}{3 \kappa ^2 \left(q^2+r^2\right)}-\frac{1}{3 \kappa ^2 \left(q^2+r^2\right)^{3/2}}
    \nonumber\\
&&       + \frac{M \left(30 q^6+15 q^4 r^2+35 q^2 r^4+14 r^6\right)}{15 \kappa ^2 q^4 r \left(q^2+r^2\right)^{5/2}} \,.
\end{eqnarray}
These expressions satisfy all equations of motion and Eq.~\eqref{RC}.

For completeness, the trace of the field equation is given by 
\begin{eqnarray}
    \kappa^2\Theta+R = -2 \kappa ^2 \left[f_1 q^2 \left(q^2+3 r^2\right)-2 f_0\right].
    \label{scalars2_BH3}
\end{eqnarray}
For this case the Kretschmann scalar is 
\begin{eqnarray}
    K&=&\frac{M^2 \left(4 q^8+44 q^6 r^2+117 q^4 r^4-158 q^2 r^6+81 r^8\right)}{r^6 \left(q^2+r^2\right)^5}\nonumber\\
    &&+\frac{8 M \left(2 r^2 \left(\sqrt{q^2+r^2}-2\right)+q^2\right)}{r \left(q^2+r^2\right)^4}\nonumber\\
    &&+\frac{6-8 \sqrt{q^2+r^2}+4 q^2+4 r^2}{\left(q^2+r^2\right)^3}.
\end{eqnarray}
which diverges in the limit $r\rightarrow 0$, but assumes an asymptotically flat behavior for $r\rightarrow \infty$.

\begin{figure}[!h]
\centering
\includegraphics[scale=0.45]{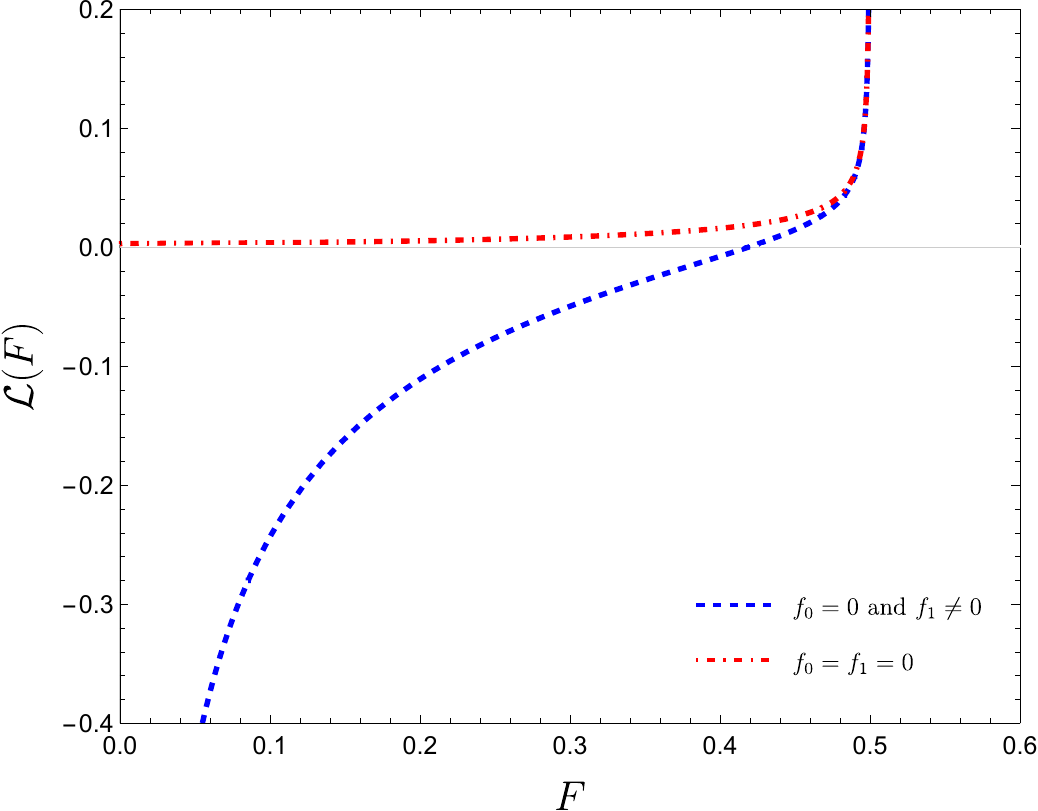}
\caption{Graphical representation of the expression ${\cal L}(F) $, described by Eq.~\eqref{L3_BH3}. The blue dashed curve represents the behavior of ${\cal L}(F)$ versus $F$ with $f_0=0$ and $f_1=0.2$, while the red dotted-dashed curve illustrates the behavior of ${\cal L}(F)$ versus $F$ with $f_0=f_1=0$. The values of the constants used are $q=1$, $\kappa=8\pi$ and $M=2$. } 
\label{LxF_BH3}
\end{figure}

As before, one may deduce the form of $r(F)$ and $r(\varphi)$, so that ${\cal L}(F)$ and $V(\varphi)$ are given by
\begin{align}
 {\cal L}(F) &= f_1 \left(q^4-\frac{q^3}{\sqrt{2} \sqrt{F}}\right)+\frac{3 \sqrt{2} \sqrt{F} q-2^{3/4} F \sqrt{\frac{q}{\sqrt{F}}}}{3 \kappa ^2 q^2}\nonumber\\
 &+f_0+\frac{7 \sqrt[4]{2} M \sqrt{q \left(\frac{\sqrt{2}}{\sqrt{F}}-2 q\right)} \left(\sqrt{2} \sqrt{F} q-2\right)}{15 \kappa ^2 q^4 \sqrt{\frac{q}{\sqrt{F}}} \left(2 \sqrt{F} q-\sqrt{2}\right)}
  \nonumber\\
& +\frac{2 \sqrt[4]{2} F M \sqrt{q \left(\frac{\sqrt{2}}{\sqrt{F}}-2 q\right)} \left(13-36 \sqrt{2} \sqrt{F} q\right)}{15 \kappa ^2 q^2 \sqrt{\frac{q}{\sqrt{F}}} \left(2 \sqrt{F} q-\sqrt{2}\right)}
.\label{L3_BH3}
\hspace{-1cm}
\\
V(\varphi)&=-\frac{7 M \left(e^{2 \sqrt{2} \kappa  \sqrt{\epsilon } \varphi }-q^2\right)^{3/2} \left(3 q^2+2 e^{2 \sqrt{2} \kappa  \sqrt{\epsilon } \varphi }\right)}{15 \kappa ^2 q^4 \left(e^{2 \sqrt{2} \kappa  \sqrt{\epsilon } \varphi }\right)^{5/2}}\nonumber\\
&-\frac{1}{6 \kappa ^2 \left(e^{2 \sqrt{2} \kappa  \sqrt{\epsilon } \varphi }\right)^{3/2}}-\frac{2 M \sqrt{e^{2 \sqrt{2} \kappa  \sqrt{\epsilon } \varphi }-q^2}}{\kappa ^2 \left(e^{2 \sqrt{2} \kappa  \sqrt{\epsilon } \varphi }\right)^{5/2}}.
\hspace{-1cm}
\label{varphi_BN3}
\end{align}
Figure~\ref{LxF_BH3} depicts the behavior of Eq.~\eqref{L3_BH3}, where the blue dashed curve represents the values of $f_0=0$ and $f_1=0.2$, while the red dotted-dashed curve illustrates the behavior at $f_0=0$ and $f_1=0$. The latter represents the behavior of GR. 

Given that $F=q^{2}/2\left(q^{2}+r^{2}\right)^{2}$, we observe that the minimum value for  $F$ occurs when $r=0$. Thus, the resulting Lagrangian Eq.~\eqref{L3_BH3}, approximately, in the limit $F\gg 1$, is: 
\begin{equation}
   {\cal L}(F) \sim f_0+f_1 q^4+\frac{24\sqrt[4]{2}M\,F^{5/4}}{5\kappa^{2}q^{3/2}}.
\end{equation}
However, in the limit $F\ll1$, a discrepancy between the two theories becomes clear. While the curve described by CKG leads to a divergent Lagrangian in this limit, the GR Lagrangian tends towards zero
\begin{equation}
   {\cal L}(F) \sim \frac{15 f_0 \kappa ^2 q^4+15 f_1 \kappa ^2 q^8+14 M}{15 \kappa ^2 q^4}-\frac{f_1 q^3}{ \sqrt{2F}}.
\end{equation}
\par

\begin{figure}[!h]
\centering
\includegraphics[scale=0.45]{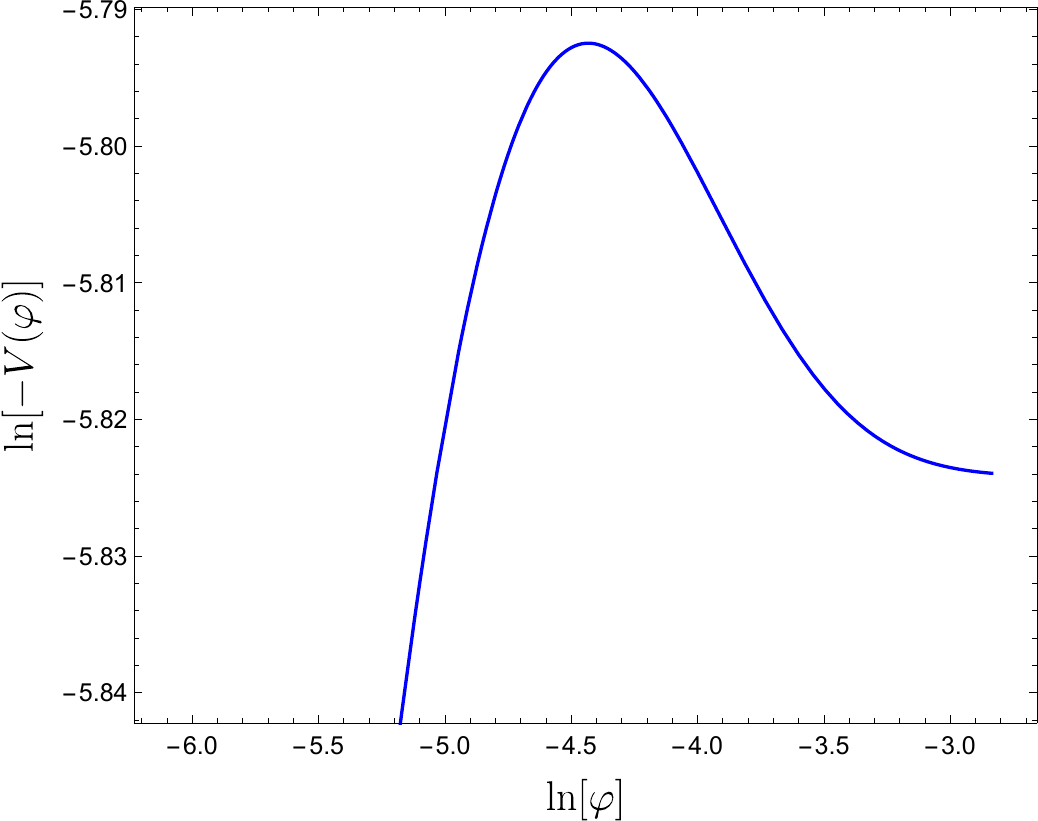}
\caption{Graphical representation of  $\ln[-V(\varphi)]$, described by Eq.~\eqref{varphi_BN3}. Where we consider $\kappa=8\pi$, $M=2.0$, $\epsilon =1.0$ and $q=0.1$ .} 
\label{Vxphi_BH2}
\end{figure}

Figure~\ref{Vxphi_BH2} illustrates the behavior of Eq.~\eqref{varphi_BN3} with respect to $\varphi$ using a parametric graph, where we have used the logarithmic function in this expression for a more detailed analysis.

\section{Regular black hole solutions} \label{sec4}

In this section,  for simplicity, will assume the absence of a scalar field, i.e.,  $V(\varphi)=0$ and $\varphi=0$. We also consider that $\Sigma(r)=r$, so that the metric
\begin{equation}
ds^2=e^{a(r)}dt^2-e^{b(r)}dr^2-r^2\left[d\theta^{2}+\sin^{2}\left(\theta\right)d\phi^{2}\right]\,,\label{m_RBH}
\end{equation} 
is used to solve the field equations \eqref{eq_CKG} in order to determine ${\cal L} {}_{\rm NED }$ and $\cal{L} _F$, which in this case coincide with Eqs. \eqref{L_BH2} and ~\eqref{LF_BH2}. 

These quantities satisfy the equations of motion, except for a single component ``1,1,1",  namely,  Eq. (\ref{111}), which we reproduce here for self-completeness:
\begin{eqnarray}
-r\left[a''(r)+b''(r)\right]+a'(r)\left[2rb'(r)+1\right]
	\nonumber \\
+ra'(r)^{2}+rb'(r)^{2}+b'(r)=0.
\end{eqnarray}
In this case, we take the functional form of $b(r)$ as follows:
\begin{equation}
    b(r)=	\begin{cases}
-a(r),\\
-a(r)-\ln\left(b_{0}+r^{2}\right)\,.
\end{cases}
    \label{b_Bardeen}
\end{equation}
Where $b_0$ and is a constant. 

\subsubsection{Bardeen-type solutions: Model with $b(r)=-a(r)$}\label{sub1_RBH}

In this model we will consider the specific solution $b(r)=-a(r)$, which satisfies all of the  equations of motion. 
To proceed with our calculations, we will use the following metric function:
\begin{equation}
e^{a(r)}=e^{-b(r)}= \left(1-\frac{2 M r^2}{\left(q^2+r^2\right)^{3/2}}-\frac{\Lambda  r^2}{3}-\frac{\lambda  r^4}{5}\right).\label{a_Bardeen}
\end{equation}
Thus, the electromagnetic quantities are readily given by:
\begin{align}
 {\cal L}_{\rm NED} (r) &= f_0+q^2 \left(\frac{6 M}{\kappa ^2 \left(q^2+r^2\right)^{5/2}}-f_1 r^2\right)\,,
 	\label{L2_Bardeen}\\
 {\cal L}_F (r) &= r^6 \left(f_1+\frac{15 M}{\kappa ^2 \left(q^2+r^2\right)^{7/2}}\right)\,,
 	\label{LF2_Bardeen}
\end{align}
which satisfy condition \eqref{RC}.

The Kretschmann scalar is now
\begin{eqnarray}
    K&=& \frac{12 M^2 \left(8 q^8-4 q^6 r^2+47 q^4 r^4-12 q^2 r^6+4 r^8\right)}{\left(q^2+r^2\right)^7}
	\nonumber \\    
   && \hspace{-0.55cm} +\frac{8 M \biggl[10 q^4 \left(2 \Lambda +3 \lambda  r^2\right)-q^2 \left(54 \lambda  r^4+5 \Lambda  r^2\right)+6 \lambda  r^6\biggr]}{5 \left(q^2+r^2\right)^{7/2}}
   \nonumber\\
 && \phantom{k=}  +\frac{8 \Lambda ^2}{3}+\frac{212 \lambda ^2 r^4}{25}+8 \lambda  \Lambda  r^2
 \,,
 \label{K_Bardeen}
\end{eqnarray}
which is regular at the origin $r\rightarrow0$. For the specific case of $\lambda\rightarrow 0$ the scalar also has a regular behavior in the limit $r\rightarrow\infty$.

Taking into account the trace of the field equation
\begin{equation}
    \kappa^2\Theta+R=4 f_0 \kappa ^2-6 f_1 \kappa ^2 q^2 r^2-4 \Lambda -6 \lambda  r^2\,,
    \label{scalars_Bardeen}
\end{equation}
where the specific case of $f_0=f_1=\lambda=0$ reduces to GR.

As before, one can determine $r(F)$, so that the electromagnetic Lagrangian is given by
\begin{equation}
    {\cal L}(F)=f_0+q^2 \left(\frac{6 M}{\kappa ^2 \left(\frac{q}{ \sqrt{2 F}}+q^2\right)^{5/2}}-\frac{f_1 q}{ \sqrt{2 F}}\right).\label{L2_RBH}
\end{equation} 
Thus all equations of motion are satisfied, as is the relation~\eqref{RC}, for this Lagrangian. We verify here that the matter is more general than in the case of GR, with an added term in the Lagrangian, and note that for $f_1=0$ we have the Bardeen Lagrangian. 

\begin{figure}[!h]
\centering
\includegraphics[scale=0.47]{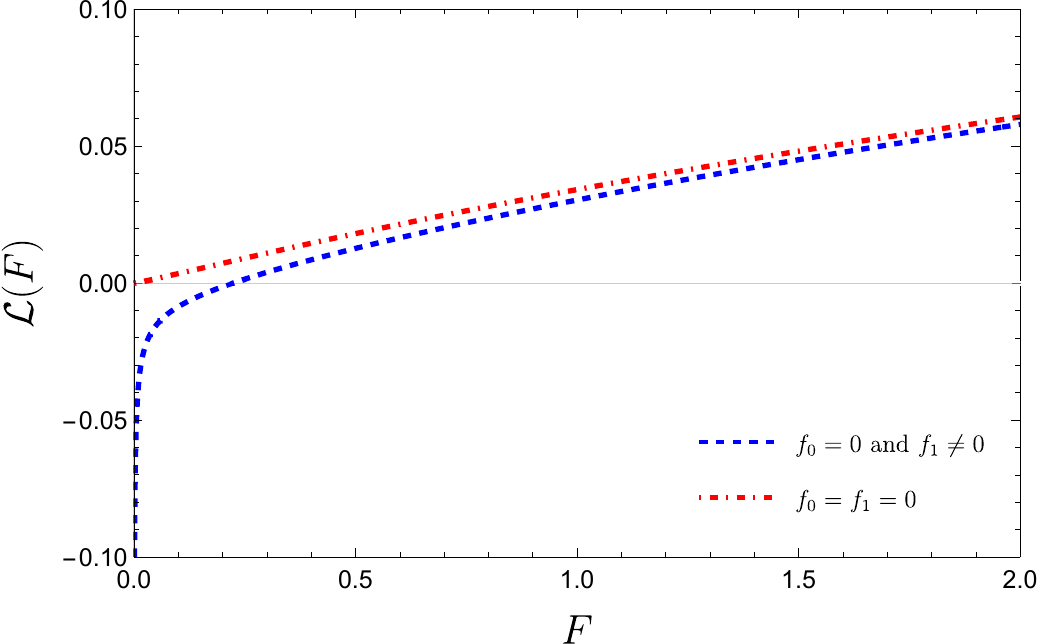}
\caption{The plot representsf the Lagrangian ${\cal L}(F) $, given by Eq.~\eqref{L2_RBH}. The blue dashed curve represents the behavior of ${\cal L}(F)$ versus $F$ with $f_0=0$ and $f_1=0.2$, while the red dotted-dashed curve illustrates the behavior of ${\cal L}(F)$ versus $F$ with $f_0=f_1=0$. The values of the parameters used are  $q=0.3$, $\kappa=8\pi$ and $M=2.0$. } 
\label{LxF_RBH}
\end{figure}

Figure \eqref{LxF_RBH} depicts the behavior of Eq.~\eqref{L2_RBH}, where the blue dashed curve represents the values of $f_0=0$ and $f_1=0.2$, while the red dotted-dashed curve illustrates the behavior with $f_0=0$ and $f_1=0$. 

If we analyze the limit $F\gg 1$ in the Lagrangian described by Eq.~\eqref{L2_RBH}, we find that the Lagrangian ${\cal L}(F)$ of the CKG theory has a remarkably similar behavior to that of GR, as shown by the curves in Fig.~\ref{LxF_RBH}. So we have  
\begin{equation}
    {\cal L}(F)\sim f_{0}+\frac{6Mq^{2}}{\kappa^{2}\left(\frac{q}{\sqrt{2F}}+q^{2}\right)^{5/2}}.
\end{equation} 
However, for the limit $F\ll 1$, a discrepancy between the two theories appears. While the CKG curve leads to a divergent Lagrangian in this scenario, the Lagrangian tends to zero in the context of GR. The approximate Lagrangian is now 
\begin{equation}
    {\cal L}(F)\sim f_{0}-\frac{f_{1}q^{3}}{\sqrt{2F}}.
\end{equation}

\par

\subsubsection{Bardeen-type solutions: Model with $b(r)=-a(r)-\ln\left(b_{0}+r^{2}\right)$}\label{sub2_RBH}

In this second case, the metric functions have the following form:
\begin{equation}
    e^{a(r)}=\left(1-\frac{2 M r^2}{\left(q^2+r^2\right)^{3/2}}-\frac{\lambda  r^4}{5}-\frac{\Lambda  r^2}{3}\right) \,,
    \label{expa_Bardeen}
\end{equation}
and 
\begin{equation}
    e^{-b(r)}=\left(b_0+r^2\right) \left(1-\frac{2 M r^2}{\left(q^2+r^2\right)^{3/2}}-\frac{\lambda  r^4}{5}-\frac{\Lambda  r^2}{3}\right).
    \label{expb_Bardeen}
\end{equation}
\par

With the above conditions, the functions ${\cal L}_{\rm NED } (r)$ and ${\cal L}_{F} (r)$ are given by
\begin{eqnarray}
 {\cal L}_{\rm NED} (r) &=&  \frac{Mr^{2}\left(-15b_{0}q^{2}+4q^{4}-10q^{2}r^{2}+r^{4}\right)}{\kappa^{2}\left(q^{2}+r^{2}\right)^{7/2}}
	\nonumber \\ 
 && \hspace{-0.85cm} +\frac{3M\left(2b_{0}q^{4}+7b_{0}q^{2}r^{2}+4q^{2}r^{4}-r^{6}\right)}{\kappa^{2}\left(q^{2}+r^{2}\right)^{7/2}} 
 	\nonumber\\
 &&  \hspace{-0.85cm}  +\frac{b_{0}-1}{\kappa^{2}r^{2}}+\frac{2-2b_{0}}{\kappa^{2}r^{2}}+f_{0}-f_{1}q^{2}r^{2}+\frac{4\lambda r^{4}}{5\kappa^{2}},\label{L3_Bardeen}
 \end{eqnarray}
 \begin{eqnarray}
 {\cal L}_F (r) &=& \frac{r^{6}}{5\kappa^{2}q^{2}}\Bigg[-\frac{5M\left(-15b_{0}q^{2}+4q^{4}-10q^{2}r^{2}+r^{4}\right)}{\left(q^{2}+r^{2}\right)^{7/2}}
	\nonumber \\ 
	&& +\frac{5-5b_{0}}{r^{4}}-8\lambda r^{2}\Bigg]+f_{1}r^{6}\,,
	\label{LF3_Bardeen}
\end{eqnarray}
which satisfy the condition~\eqref{RC}. 

In this case, the Kretschmann scalar in regular in the limits of $r\rightarrow0$ and $r\rightarrow\infty$, so that spacetime is regular,  for $\lambda\rightarrow0$. 
Consider now the trace of the field equation:
\begin{eqnarray}
    \kappa^2\Theta +R &=& -4 b_0 \Lambda -6 b_0 \lambda  r^2+4 f_0 \kappa ^2-6 f_1 \kappa ^2 q^2 r^2
	\nonumber \\    
    &&-\frac{12 M r^2}{\left(q^2+r^2\right)^{3/2}}-\frac{1}{5} 6 \lambda  r^4-6 \Lambda  r^2+6\,,
    \label{scalars2_Bardeen}
\end{eqnarray}
which does not reduce to GR.

Now by writing $r(F)$, we can express
\begin{eqnarray}
{\cal L}(F)&=& \frac{3Mq^{3}\left(8b_{0}F^{3/2}q+14\sqrt{2}b_{0}F+8\sqrt{F}q-\sqrt{2}\right)}{4F^{3/2}\kappa^{2}\left(\frac{q}{\sqrt{2}\sqrt{F}}+q^{2}\right)^{7/2}}
	\nonumber \\
	&& +\frac{4M\left(-30b_{0}F+8Fq^{2}-10\sqrt{2}\sqrt{F}q+1\right)}{\kappa^{2}\sqrt{q\left(\frac{\sqrt{2}}{\sqrt{F}}+2q\right)}\left(2\sqrt{F}q+\sqrt{2}\right)^{3}}
	\nonumber\\	
&& +f_{0}-\frac{f_{1}q^{3}}{\sqrt{2 F}}-\frac{\sqrt{2}(b_{0}-1)\sqrt{F}}{\kappa^{2}q}+\frac{2\lambda q^{2}}{5F\kappa^{2}}\,,
\label{L4_Bardeen}
\end{eqnarray}
which satisfies all the equations of motion and the condition~\eqref{RC}. Here, as this case does not reduce to GR, we cannot use Eq. \eqref{L4_Bardeen} to find the Bardeen Lagrangian as a particular case. 

\begin{figure}[h]
\centering
\includegraphics[scale=0.45]{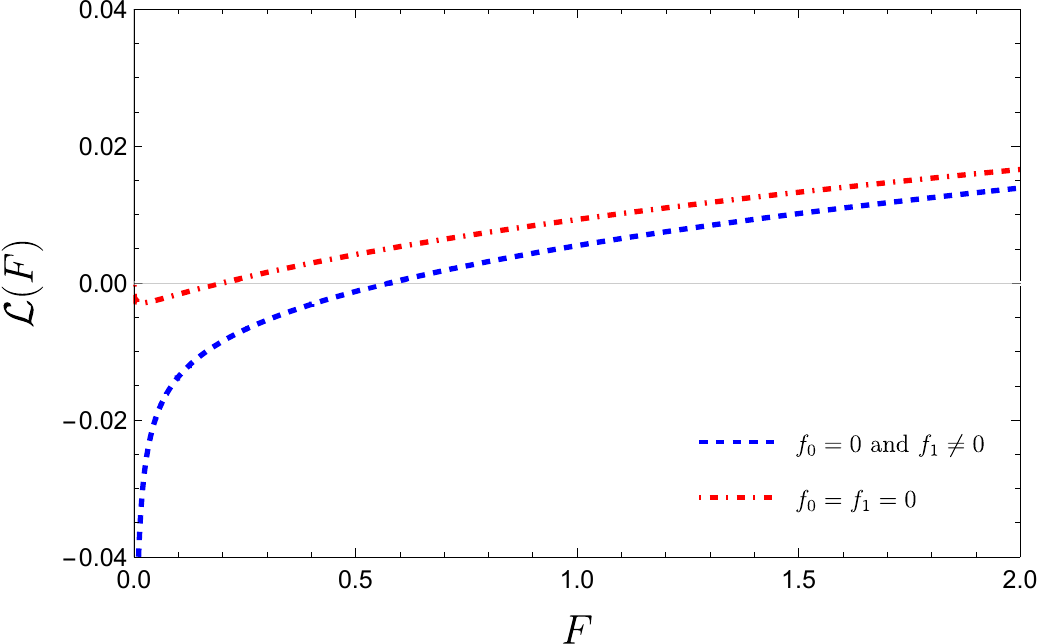}
\caption{Graphical representation of the expression ${\cal L}(F) $, described by Eq.~\eqref{L4_Bardeen}. The blue dashed curve represents the behavior of ${\cal L}(F)$ versus $F$ with $f_0=0$ and $f_1=0.2$, while the red dot-dashed curve illustrates the behavior of ${\cal L}(F)$ versus $F$ with $f_0=f_1=0$. The values of the constants used are  $\lambda=0$, $k_0=0.4$, $b_0=0.1$,  $q=0.3$, $\kappa=8\pi$ and $M=2.0$ } 
\label{LxF2_RBH}
\end{figure}

Figure \eqref{LxF2_RBH} depicts the behavior of the Eq.~\eqref{L4_Bardeen} by the blue dashed curve for values of $f_0=0$ and $f_1=0.2$, and by the red dotted-dashed curve for $f_0=0$ and $f_1=0$. Similar to the previous model, the limit for $F\gg1$ in the Lagrangian described by Eq.~\eqref{L4_Bardeen} exhibits a behavior consistent with GR, as shown in Fig.~\ref{LxF2_RBH}. In this limit, we can express an approximation for  ${\cal L}(F)$ as follows:
\begin{equation}
   {\cal L}(F)\sim\left(\frac{6 b_0 M q^4}{\kappa ^2 \left(q^2\right)^{7/2}}+f_0\right)-\frac{\sqrt{2} (b_0-1) \sqrt{F}}{\kappa ^2 q}.  
\end{equation}
The distinction remains when the limit $F\ll1$ is examined, but the curve representing the GR Lagrangian now converges to a fixed negative value. So now we have the following approximation for this limit
\begin{equation}
   {\cal L}(F)\sim f_0-\frac{f_1 q^3}{\sqrt{2} \sqrt{F}}.  
\end{equation}

\par

\section{Summary and Discussion}\label{sec:concl}

In this work, we explored black hole and regular black hole solutions in the recently proposed Conformal Killing gravitational theory. This theory is of third order in the derivatives of the metric tensor and essentially satisfies three theoretical criteria for gravitational theories beyond GR. These criteria essentially stipulate the following: (i) obtain the cosmological constant as an integration constant; (ii) derive the energy conservation law as a consequence of the field equations, rather than assuming it; (iii) and not necessarily consider conformally flat metrics as vacuum solutions.  In fact, existing modified theories of gravity, including GR, do not simultaneously fulfil all of these three criteria.  Here, we coupled CKG to scalar fields and NLED, and we explored solutions of black holes and regular black holes. We found generalizations of the Schwarschild--Reissner-Nordstr\"{o}m--Ads solutions, and further extended the class of regular black hole solutions found in the literature.

In this work, we analyzed possible solutions for magnetically charged black holes. Among them is a solution described by the metric~\eqref{a_Maxwell}, where the constant $k_0$ should be identified as $q^2$. The final solution is obtained as an extension of the Reissner-Nordström solution with a linear ${\cal L}(F)$. 

In the second solution examined, where we considered $\varphi=V(\varphi)=0$, we found two possible solutions for the equations of motion. One of them is described by the metric function~\eqref{a_BH2} with a new term given by $k_0/r^4$. In this context, we investigated the possible configurations for event horizons depending on fixing the values of some parameters $M$, $q$, $\Lambda$, $\lambda$. We note in condition~\eqref{scalars_BH2} ($\kappa^2\Theta+R=4 f_0 \kappa ^2-6 f_1 \kappa ^2 q^2 r^2-4 \Lambda -6 \lambda  r^2$) that we attain the case of general relativity when $f_0=f_1=\lambda=0$. For other values, however, the theory differs from GR and represents an alternative theory. Another important feature is that the Lagrangian for this model introduces new terms due to the inclusion of $f_1$ and $k_0$, as described by equation~\eqref{L3_BH2}.  In the limit $F\gg 1$ we found that the CKG Lagrangian expands infinitely and behaves similarly to that of the GR. In the limit $F\ll 1$, however, the theories differ considerably, with the GR Lagrangian tending to zero. 

In the second model, we considered the same metric function defined by Eq.~\eqref{a_BH2}, but with the functional form $b(r)=-a(r)- \ln(b_0+r^2)$. The trace of this case, expressed by Eq.~\eqref{scalars2_BH2}, is identical to the GR case only if we consider $f_0=f_1=\lambda=k_0=b_0=0$. We observe that in this scenario, the Lagrangian we found in Eq.~\eqref{L3_BH2} presents new terms with $b_0$ and $k_0$ that differentiate it from GR. Similar to the previous scenario, the Lagrangians of the CKG and GR theories show similar behavior at the limit $F\gg 1$. For the limit $F\ll 1$, however, there are differences and the GR Lagrangian now converges to a fixed negative value. When $r\rightarrow \infty$,we see that the Ricci scalar of this model is $R\rightarrow 6$, i.e. a non-zero constant, which shows that this region is still curved, even if the cosmological constant is zero. This can be interpreted with the metric functions, in this limit, $e^{a(r)} \rightarrow  1$ and $e^{b(r)}\sim r^2$, so that even at radial infinity we have non-zero curvature.

In the third black hole model we considered the hypothesis $b(r)=-a(r)+b_0(r)$ and the metric function is given by \eqref{a_BH2}. We note that the trace~\eqref{scalars2_BH3} introduces new terms with $f_0$ and $f_1$. However, by examining the trace of this model, we found that GR is not recovered. It is important to note that the Kretschmann scalar diverges only in the limit of $r\rightarrow 0$. These new terms are also contained in the Lagrangian of this model, which is described by the equation \eqref{L3_BH3}. This clearly shows that this theory differs from GR. Similar to previous models, the CKG Lagrangian shows a difference to GR in the limit value $F\ll 1$.

In the last model, we investigated solutions for regular black holes as described in Section~\ref{sec4}, but we returned to considering only NLED as the source of matter. In this approach, we adopted a functional form for $b(r)$ that provides two possible solutions. In this way, we obtained two different models, both using the type Bardeen metric function~\eqref{a_Bardeen}.
For the first solution with $b(r)=-a(r)$, we note in trace~\eqref{scalars_Bardeen} that we only return to the case of general relativity if $f_0 = f_1 = \lambda = 0$. We also note that the Lagrangian term contains new terms that distinguish it from the GR. In the limit of $F\gg 1$ we find that the Lagrangian of CKG behaves similarly to that of GR, but in the limit of $F\ll 1$ the theories differ considerably. In the second model, we  assumed $b(r)=-a(r)-\ln\left(b_{0}+r^{2}\right)$ to obtain the solutions. In this scenario, we can see from the trace that we cannot recover GR, and this is already clear when we find the metric function Eq.~\eqref{expb_Bardeen}. We verified that the theories are close to each other in the limit $F\gg1$, but differ significantly in the limit $F\ll1$. 
Therefore, upon examining all the solutions presented in this manuscript, we have found that the Lagrangians are consistently nonlinear and do not recover the solutions of GR. If we analyze the Ricci scalar for this model with $\lambda=\Lambda=0$, we find that the Ricci scalar becomes $R\rightarrow 6$ when we consider the limit $r\rightarrow\infty$, a non-zero constant. This result shows that the region in question still exhibits curvature. This interpretation is supported by the metric functions that occur in this limit $a(r)\rightarrow1$ and $b(r)\sim r^2$. So even at radial infinity, the curvature is not zero. 

With the exception of the first model presented, all other solutions do not reduce to the Maxwell Lagrangian, consistently remaining nonlinear. This is exemplified, for instance, in the solution proposed by Ayon-Beato Garcia for the Bardeen regular  black hole~\cite{Ayon-Beato:2000mjt}.

The analysis carried out in this work has further extended the class of (regular) black hole solutions explored in the literature.  In the future, we plan to investigate several other approaches within CKG, that include comprehensive studies of black hole thermodynamics, stability of solutions through perturbation analysis, detailed studies of black hole shadows, and gravitational lensing analysis. These are just some of the areas we plan to address in our future work to deepen our understanding and further contribute to the development of this new alternative approach to GR.

Since the metric functions can be the same for both GR and CKG and only the material content changes, where the latter being more general in the case of CKG and consisting of one or more terms, it would be important to verify the existence and uniqueness of the solutions presented in this work. However, here we were only concerned with the electromagnetic material content with magnetic charge. An important question that arises is whether it would be possible to represent the same solution with electric charge, as it is done in GR.


{\bf Data availability statement}.

No new data were created or analysed in this study.

\acknowledgments{
MER thanks CNPq for partial financial support.  This study was supported in part by the Coordenção de Aperfeioamento de Pessoal de Nível Superior - Brazil (CAPES) - Financial Code 001.  
FSNL acknowledges support from the Funda\c{c}\~{a}o para a Ci\^{e}ncia e a Tecnologia (FCT) Scientific Employment Stimulus contract with reference CEECINST/00032/2018, and funding through the research grants UIDB/04434/2020, UIDP/04434/2020, CERN/FIS-PAR/0037/2019 and PTDC/FIS-AST/0054/2021.}


%

\end{document}